\documentclass[12pt]{article}
\usepackage{amsmath,amssymb}
\renewcommand{\theequation}{\arabic{section}.\arabic{equation}}

\font\onefour=cmr10 at 16pt

\newcommand{\vTs}{\vphantom{\mbox{\onefour I}}}

\begin{document}



\def\a{{\bf a}}
\def\b{{\bf b}}
\def\d{\delta}
\def\e{{\bf e}}
\def\f{{\bf f}}
\def\g{{\bf g}}
\def\h{\mathfrak{h}}
\def\i{{\rm i}}
\def\k{\kappa}
\def\l{\lambda}
\def\o{\omega}
\def\p{\wp}
\def\r{\rho}
\def\t{{\bf t}}
\def\s{{\rm str}}
\def\u{{\bf u}}
\def\v{{\bf v}}
\def\z{{\bf z}}
\def\x{\xi}
  \def\A{{\cal{A}}}
  \def\B{{{\bf B}}}
  \def\C{{\cal{C}}}
  \def\D{{\cal{D}}}
\def\F{{\bf F}}
\def\G{\Gamma}
\def\K{{\cal{K}}}
\def\O{\Omega}
\def\R{\bar{R}}
\def\T{{\cal{T}}}
\def\V{{\bf V}}
\def\L{\Lambda}
\def\X{{\bf X}}
\def\Zb{\mathbb{Z}}
\def\Cb{\mathbb{C}}

\def\R{\overline{R}}

\def\beq{\begin{equation}}
\def\eeq{\end{equation}}
\def\bea{\begin{eqnarray}}
\def\eea{\end{eqnarray}}
\def\ba{\begin{array}}
\def\ea{\end{array}}
\def\no{\nonumber}
\def\le{\langle}
\def\re{\rangle}
\def\lt{\left}
\def\rt{\right}

\newtheorem{Theorem}{Theorem}
\newtheorem{Definition}{Definition}
\newtheorem{Proposition}{Proposition}
\newtheorem{Lemma}{Lemma}
\newtheorem{Corollary}{Corollary}
\newcommand{\proof}[1]{{\bf Proof. }
         #1\begin{flushright}$\Box$\end{flushright}}

\baselineskip=20pt

\newfont{\elevenmib}{cmmib10 scaled\magstep1} \newcommand{\preprint}{
    \begin{flushleft}
      \elevenmib Yukawa\, Institute\, Kyoto\\
    \end{flushleft}\vspace{-1.3cm}
    \begin{flushright}\normalsize
    \sf  YITP-06-64\\
      {\tt hep-th/0612154} \\ December 2006
    \end{flushright}}
\newcommand{\Title}[1]{{\baselineskip=26pt
    \begin{center} \Large \bf #1 \\ \ \\ \end{center}}} \newcommand{\Author}{\begin{center}
    \large \bf
Ryu Sasaki,${}^a$ ~Wen-Li Yang${\,}^{b,c}$~ and~Yao-Zhong
Zhang${\,}^{a,b}$\end{center}}
\newcommand{\Address}{\begin{center}

      ${}^a$ Yukawa Institute for Theoretical Physics, Kyoto
      University, Kyoto 606-8502, Japan\\
      ${}^b$ Department of Mathematics, University of Queensland, Brisbane, QLD 4072,
      Australia\\
  ${}^c$ Institute of Modern Physics, Northwest University,
      Xian 710069, P.R. China
    \end{center}}
\newcommand{\Accepted}[1]{\begin{center}
    {\large \sf #1}\\ \vspace{1mm}{\small \sf Accepted for Publication}
    \end{center}}

\preprint \thispagestyle{empty}
\bigskip\bigskip\bigskip

\Title{Exact classical solutions of   nonlinear sigma models on
supermanifolds} \Author

\Address \vspace{1cm}

\begin{abstract}
We study   two-dimensional nonlinear sigma models with target
spaces being the complex super Grassmannian manifolds, that is,
coset supermanifolds $G(m,p|n,q)\cong U(m|n)/[U(p|q)\otimes
U(m-p|n-q)]$ for $0\leq p \leq m$, $0\leq q \leq n$ and $1\leq
p+q$. The projective superspace ${\bf CP}^{m-1|n}$ is  a special
case of $p=1$, $q=0$. For the two-dimensional Euclidean base
space, a wide class of exact classical solutions (or harmonic
maps) are constructed explicitly and elementarily in terms of
Gramm-Schmidt orthonormalisation procedure starting from
holomorphic bosonic and fermionic supervector input functions. The
construction is a  generalisation of the non-super case published
more than twenty years ago by one of the present authors.

\vspace{1truecm}

\end{abstract}

\newpage
\section{Introduction}
\label{intro} \setcounter{equation}{0}

A wide class of exact classical solutions are constructed in this
paper, for Euclidean two-dimensional non-linear sigma models on
complex super Grassmannian manifolds. The general
motivation/background of the present work is the recent interest in
2-D non-linear sigma models on supergroups, in particular
$PSU(1,1|2)$, $PSU(2,2|4)$ and more generally $PSL(n|n)$
\cite{Berk} and some of their supercoset manifolds. They are
related to superstrings propagating on certain $AdS$ backgrounds.
These models have also found applications in statistical mechanics,
such as the integer quantum Hall effect and its recent
generalisation, fermions with quenched disorder, percolation,
polymers etc \cite{others}. Like their non-super counterparts,
these 2-D sigma models are classically integrable and enjoy an
infinite number of local/non-local conservation laws \cite{cons}.
In contrast to the non-super 2-D sigma models for which masses are
dynamically generated by quantum effects, exact conformal
invariance is preserved at the quantum level in some special
supergroup sigma models \cite{Berk}. This would mean, as in higher
$\mathcal{N}$ supersymmetric gauge theories, that the quantum
theory and classical theory are closely related and that quantum
results could be inferred/derived from their relatively
well-understood classical counterparts. In fact, various integrable
structures and methods, for example, an infinite number of
local/non-local conserved quantities, symmetry transformations,
etc., have much simpler forms at the classical level than the
quantum ones. A naive hope arises that exact and fairly general
classical solutions, if available, could elucidate various aspects
of the corresponding quantum field theory.

The present paper is a modest attempt into that general direction
by providing a fairy wide class of exact solutions for non-linear
sigma models on certain supercosets, namely the complex super
Grassmannian. These are the supermanifold (not spacetime
supersymmetric) versions of the complex projective space ${\bf
CP}^{N-1}$ and complex Grassmannian sigma models \cite{Wit79}.
Together with the sine-Gordon theory and $O(n)$ sigma models, they
have been investigated quite extensively as a theoretical
laboratory for four dimensional gauge theories for about a quarter
of a century. For the  complex Grassmannian $G(N,m)\cong
U(N)/[U(m)\times U(N-m)]$ sigma models
  (including  the ${\bf CP}^{N-1}$ \cite{DinWoj} as a special case)
  on  2-D Euclidean space, quite general classical solutions
(or harmonic maps in mathematics)  were constructed by one of the
present authors more than twenty years ago \cite{Sas83}. Starting
from $m$    holomorphic input vector functions,  Gramm-Schmidt
orthonormalisation procedure is applied to produce $N$ unit column
vectors of the corresponding $U(N)$ group. Certain subsets of
these unit vectors constitute solutions of the complex
Grassmannian sigma model. Although the structure of the super
Grassmannian $G(m,p|n,q)\cong U(m|n)/[U(p|q)\otimes U(m-p|n-q)]$
is much more complicated than the non-super $G(N,m)\cong
U(N)/[U(m)\times U(N-m)]$, the basic strategy of solution
construction is about the same. One starts with  $p$ bosonic and
$q$ fermionic holomorphic input  supervector functions and
orthonormalise them in terms  of Gramm-Schmidt procedure to
produce $m+n$ basis vectors  of  the super unitary group $U(m|n)$.
Again certain subsets of these unit supervectors constitute
solutions of the complex super Grassmannian sigma model.

The paper is organised as follows. In section two, basic concepts
and notation for supernumbers, supervectors, supervector spaces,
supermatrices and super Grassmannians are introduced. The
two-dimensional non-linear sigma models  on the super Grassmannian
manifolds are introduced in section three. The gauge invariant
action, equations of motion in various equivalent forms are
derived and symmetry properties are explored. Exact solutions are
constructed in section four starting from  $p$ bosonic and $q$
fermionic input supervectors. Gramm-Schmidt orthonormalisation
procedure is applied to produce the  unit supervectors of the
super unitary group $U(m|n)$. The final  section is for comments
and a summary. The Appendix provides an elementary proof of one
important formula (\ref{Relation-1}) in section four.


\section{Supervector space and super Grassmannian} \label{NP}
\setcounter{equation}{0}
Let us start with a few words on the Grassmann algebra and the
supernumbers \cite{Dew85}. Here we use the standard notion of
the Grassmann algebra $\Lambda_N$ over $\mathbb{C}$, generated
by $\xi^a$, $a=1$,\ldots, $N$, which anticommute
\[
\xi^a\xi^b=-\xi^b\xi^a,\quad (\xi^a)^2=0,\quad \forall a,b.
\]
To be more precise, we use the inductive limit of $N\to\infty$,
$\Lambda_\infty$. The elements of $\Lambda_\infty$ are called
supernumbers. Every supernumber $z$ has its body and soul
\[
z=z_B+z_S,
\]
where the body ($z_B$) is the ordinary complex
number and the soul ($z_S$) vanishes when all the Grassmann
generators are put to zero, $\xi^a\to0$. A supernumber $z$  has the
inverse $z^{-1}$ if and only if its body is non-vanishing $z_B\neq
0$.

Let us fix four non-negative integers $m$, $n$, $p$ and $q$ such
that $2\leq m+n$, $0\leq p \leq m$,  $0\leq q \leq n$ and   $1\leq
p+q$. Let $\V$ be a $\Zb_2$-graded $(m+n)$-dimensional complex
vector space (or supervector space of type $(m,n)$ \cite{Dew85})
with the pure basis $\{e_i|i=1,\ldots m+n\}$. Here $m$ is the
number of even basis vectors and $n$ is the number of odd bases.
It should be emphasised that the assignment of the $\Zb_2$-grading
to each index $j$ is completely arbitrary. One can choose the
indices of even grading $\mathbb{E}$ and odd grading $\mathbb{O}$
at will:
\begin{align}
[e_j]=0\quad \mbox{for} \ j\in\mathbb{E},\quad
\#\mathbb{E}=m,\qquad [e_j]=1\quad  \mbox{for} \
j\in\mathbb{O},\quad \#\mathbb{O}=n. \label{indgrad}
\end{align}
Once the sets $\mathbb{E}$ and $\mathbb{O}$ are specified they
must be kept fixed. One simple and often used choice is
$\mathbb{E}=\{1,2,\ldots,m\}$ and $\mathbb{O}=\{m+1,\ldots,
m+n\}$. Then any supervector $\a\in\V$ is expanded in terms of the
basis, $\a=\sum_{i}e_i\,a_i$, and it can be represented by an
$(m+n)$-component column supervector,
\begin{equation}
\a=\,\left(\begin{array}{c}a_1\\\vdots\\a_{m+n}
\end{array}\rt),\quad a_j\in\Lambda_\infty,\quad j=1,\ldots, m+n.
\label{vecnotation}
\end{equation}
A supervector with definite grading is called a pure vector. There
are bosonic and fermionic pure vectors. A  bosonic  (resp.
fermionic) pure vector $\a\in\V$ has  gradings: $[a_i]=0$ for
$i\in\mathbb{E}$ and $[a_i]=1$ for $i\in\mathbb{O}$ (resp.
$[a_i]=1$ for $i\in\mathbb{E}$
  and $[a_i]=0$ for $i\in\mathbb{O}$).
In this paper we consider pure vectors only.

The hermitian conjugate of a supervector $\a\in \V$, denoted by
$\a^{\dagger}$, is an $(m+n)$-component row supervector,
\[
\a^{\dagger}=(a_1^{*},\ldots,a^{*}_{m+n}).
\]
The complex conjugation ($*$-operation)   of supernumbers  has the
following properties \cite{Fra96},
\[
(ab)^*=b^*a^*, \qquad
(a^*)^*=a, \qquad [a^*]=[a].
\]
The supervector space ${\bf V}$ can be endowed with an inner product:
${\V\times \V}\mapsto\Lambda_\infty$, denoted by
\begin{equation}
\langle{\bf \omega},\v\rangle\in
\Lambda_\infty,\quad \omega, \v\in\V.
\end{equation}
It is bilinear in following sense
\begin{align}
&\langle \omega_1a_1+\omega_2a_2,\v_1 b_1+\v_2
b_2\rangle\nonumber\\ &\quad =a_1^*\,\langle\omega_1,\v_1\rangle
b_1+a_1^*\,\langle\omega_1,\v_2\rangle b_2+
a_2^*\,\langle\omega_2,\v_1\rangle
b_1+a_2^*\,\langle\omega_2,\v_2\rangle b_2
\end{align}
for $\omega_1,\omega_2,\v_1,\v_2\in\V$ and any supernumbers $a_j$ and
$b_j$, and enjoys,
\begin{equation}
\langle\v,{\bf
\omega}\rangle=\langle{\bf \omega},\v\rangle^*.
\end{equation}
Let the pure basis $\{e_i|i=1,\ldots,m+n\}$ be orthonormal.
Then the inner product of two supervectors $\a_1$ and $\a_2$ can
be expressed as
\begin{equation}
\langle \a_1,\a_2\rangle=\a_1^{\dagger}\a_2.
\end{equation}
The norm of a supervector $\a$, denoted by $|\!|\a|\!|$, is
defined by
\begin{equation}
|\!|\a|\!|^2=\langle \a,\a\rangle=\a^{\dagger}\a,
\end{equation}
which is an even supernumber. It is positive definite so long as
the body $\a_B$ is non-vanishing. In this case we can define a
normalised (unit) vector
\begin{equation}
\u=\a/|\!|\a|\!|,\quad \a_B\neq \mathbf{0}.
\end{equation}
It is easy to see that the body of a normalised vector is
normalised, too:
\begin{equation}
\u_B=\a_B/|\!|\a_B|\!|,\quad |\!|\u_B|\!|=1.
\label{bodyn}
\end{equation}

Let us introduce a new basis by a linear combination of the old
one:
\begin{equation}
e'_j=\sum_l e_l\,U_{l\, j},\quad j=1,\ldots,m+n.
\end{equation}
By requiring that the new basis is again orthonormal,
we obtain the condition
\begin{equation}
\sum_l U^*_{l\,j}\,U_{l\,k}=\delta_{j\,k},\quad\mbox{or}\quad
U^\dagger\, U=1_{m+n},
\label{unicond}
\end{equation}
in which
$1_{m+n}$ is the $(m+n)\times(m+n)$ identity matrix. Namely $U$ is
a unitary supermatrix. Since the new basis vectors are also
required to be pure with the same grading as the old one, there
are $m$ even and $n$ odd basis vectors. The grading of $U_{l\,j}$
is constrained as
\begin{equation}
[U_{l\, j}]=[l]+[j],\quad
\mbox{mod}\ 2.
\end{equation}
That is each column supervector of $U$
\begin{equation}
U=(\u_1,\ldots,\u_{m+n}),\qquad U\in U(m|n),
\end{equation}
is either bosonic or fermionic. In other words, there are $m$ bosonic
and $n$ fermionic column supervectors satisfying the
orthonormality condition
\begin{equation}
\u^{\dagger}_j\u_k=\delta_{j\,k},\quad j,k=1,\ldots,m+n,
\end{equation}
as rephrased  from (\ref{unicond}).

Here are some  facts and notation for $(m+n)\times(m+n)$
supermatrices used in this paper. Among  the $m+n$ indices of a
supermatrix $M$, $m$ are even $[i]=0$ and $n$ are odd $[i]=1$ and
the $\Zb_2$-grading of the indices are fixed once and for all as
in (\ref{indgrad}). A supermatrix $M$ is called even, $[M]=0$, if
$[M_{(e)\,(e)}]=[M_{(o)\,(o)}]=0$ and
  $[M_{(e)\,(o)}]=[M_{(o)\,(e)}]=1$; whereas, a supermatrix $M$ is called odd,
  $[M]=1$, if $[M_{(e)\,(e)}]=[M_{(o)\,(o)}]=1$ and
  $[M_{(e)\,(o)}]=[M_{(o)\,(e)}]=0$.
Here $(e)$ stands for the indices from $\mathbb{E}$ and $(o)$ from
$\mathbb{O}$. These are called supermatrices of definite grading
and we will consider such supermatrices only in this paper.
  The supertrace of a supermatrix $M$
  is defined by \cite{Fra96}
\begin{equation}
{\rm str} \lt(M\rt)=\sum_{j=1}^{m+n}(-1)^{([j]+[M])[j]}\,M_{j\,j}.
\label{Supertrace}
\end{equation}
The hermitian conjugate of $M$, denoted by $M^{\dagger}$, is an
$(m+n)\times(m+n)$ supermatrix with the entries
\[
\lt(M^{\dagger}\rt)_{ij}=M_{ji}^*, \qquad i,j=1,\ldots,m+n.
\]
The supertrace defined by (\ref{Supertrace}) enjoys  the following
properties,
\begin{eqnarray}
{\rm str}\lt(M^{\dagger}\rt)&=&{\rm
str}\lt(M\rt)^*, \label{Property-1}\\ {\rm
str}\lt(M_1M_2\rt)&=&(-1)^{[M_1][M_2]}\, {\rm str}\lt(M_2M_1\rt).
\label{Property-2}
\end{eqnarray}

\bigskip
The ordinary complex Grassmannian manifold $G(N,p)$ is a
collection of $p$-dimensional sub-vector spaces within a complex
$N$-dimensional vector space $\mathbb{C}^N$. A point in $G(N,p)$
is specified by a choice of $p$-orthonormal  basis vectors
$\{e^{''}_j\}$, $j=1$, \ldots, $p$ which is a subset of an
orthonormal basis $\{e'_j\}$, $j=1$, \ldots, $N$ obtained by an
arbitrary unitary transformation ($U(N)$) from a fixed orthonormal
basis $\{e_j\}$, $j=1$, \ldots, $N$. Any unitary transformations
among the chosen vectors $\{e^{''}_j\}$, $j=1$, \ldots, $p$,
($U(p)$) and among the not-chosen vectors
  $\{e^{''}_j\}$, $j=p+1$, \ldots, $N$, ($U(N-p)$) are immaterial.
Thus we have
\begin{equation}
G(N,p)=\frac{U(N)}{U(p)\times
U(N-p)}.
\label{ordCoset}
\end{equation}

In this paper we discuss the complex Grassmannian supermanifold
$G(m,p|n,q)$, which consists of a collection of sub-supervector
spaces of $(p,q)$ type within a complex supervector space $\V$ of
$(m,n)$ type. A point in $G(m,p|n,q)$ is specified by a choice of
$(p+q)$-orthonormal basis vectors $\{e^{''}_j\}$, $j=1$, \ldots,
$p+q$ among which $p$ are even and $q$ are odd. It is a subset of
an orthonormal basis $\{e'_j\}$, $j=1$, \ldots, $m+n$ obtained by
an arbitrary super unitary transformation ($U(m|n)$) from a fixed
orthonormal  basis $\{e_j\}$, $j=1$, \ldots, $m+n$. Any super
unitary transformations among the chosen vectors $\{e^{''}_j\}$,
$j=1$, \ldots, $p+q$, ($U(p|q)$) and among the not-chosen vectors
  $\{e^{''}_j\}$, $j=p+q+1$, \ldots, $m+n$, ($U(m-p|n-q)$) are immaterial.
Thus we have
\begin{equation}
G(m,p|n,q)=\frac{U(m|n)}{U(p|q)\times U(m-p|n-q)}.
\label{Coset}
\end{equation}
It is a Riemannian symmetric superspace \cite{Zir96}, as the
ordinary $G(N,p)$ is a Riemannian symmetric space. It should be
stressed that the $\Zb_2$-grading of the new chosen basis
$\{e^{''}_j\}$, $j=1$, \ldots, $p+q$, is completely independent of
the original basis  $\{e_j\}$, $j=1$, \ldots, $m+n$
(\ref{indgrad}), since it refers to the  $\Zb_2$-grading of the
new sub-supervector space of $(p,q)$ type. A different choice of
$\{e^{''}_j\}$, $j=1$, \ldots, $p+q$ with a different
$\Zb_2$-grading corresponds to a different sub-supervector space.


\section{Nonlinear sigma models on Supermanifolds}
  \label{MD} \setcounter{equation}{0}
We shall study the two-dimensional nonlinear sigma model with the
target space being this particular Riemannian symmetric superspace
$G(m,p|n,q)$. The base space is the two-dimensional Euclidean
space. The resulting sigma models are the supermanifold version of
the ordinary $G(N,p)$ models considered in \cite{Wit79,Sas83}.

Let $x=(x_1,x_2)\in\mathbb{R}^2$ be the coordinates of the
two-dimensional Euclidean space and $g=g(x)$ be a field which
takes value in $U(m|n)$. We decompose it into two  parts
\begin{equation}
g=(X,Y),\label{Fields}
\end{equation}
with
\begin{equation}
X=(\z_1,\ldots,\z_{p+q}),\qquad Y=(\z_{p+q+1},\ldots,\z_{m+n}).
\label{XY}
\end{equation}
Here $\z_i$ is an $(m+n)$-component column supervector, either
bosonic or fermionic. There are $p$ bosonic and $q$ fermionic
column supervectors in $X$ and $m-p$ bosonic and $n-q$ fermionic
column supervectors in $Y$. As explained above, the $G(m,p|n,q)$
sigma model is described by the dynamical variable $X=X(x)$,
satisfying the constraint
\begin{equation}
X^\dagger X=I_{p+q},
\label{Orthnormal}
\end{equation}
originating from the unitarity
(\ref{unicond}) of $g$. The Grassmannian structure of $G(m,p|n,q)$
is incorporated through the covariant derivative for $X$,
\begin{equation}
D_{\mu}\,X=\partial_{\mu}X-XA_{\mu}, \qquad
\mu=1,2,
\end{equation}
where the gauge potential $A_{\mu}$ is
given by
\begin{equation}
A_{\mu}=X^{\dagger}\partial_{\mu}X,\qquad \mu=1,2.
\label{Gauge-1}
\end{equation}
The constraints (\ref{Orthnormal}) imply that the gauge potential
satisfies
\begin{equation}
\lt(A_{\mu}\rt)^{\dagger}=-A_{\mu},\qquad \mu=1,2.
\label{Gauge-2}
\end{equation}
The action of the  $G(m,p|n,q)$ nonlinear
sigma model  in two-dimensional Euclidean space  is given by
\begin{equation}
S=\int d^2x\,
\s\lt(\lt(D_{\mu}X\rt)^{\dagger}\,\lt(D_{\mu}X\rt)\rt),
\label{Action}
\end{equation}
where as usual the repeated indices mean the summation.
This action has the $U(p|q)$ local gauge symmetry,
\begin{equation}
X(x)\longrightarrow X'(x)=X(x)\,h(x),\quad h(x)\in U(p|q),
\label{Gauge-tr}
\end{equation}
as well as the global $U(m|n)$ symmetry
\begin{equation}
X(x)\longrightarrow X'(x)=g_0\,X(x),\quad g_0\in U(m|n),
\quad \partial_{\mu}g_0=0.
\end{equation}
This is because the matrices $X$, $h(x)$ and $g_0$ are always even
supermatrices, i.e., $[X]=[h(x)]=[g_0]=0$,
  and the supertrace formula (\ref{Property-2})
applies without the extra sign.
The classical equation of motion of the model is
\begin{equation}
D_{\mu}D_{\mu}X+ X\,\lt(D_{\mu}X\rt)^{\dagger}D_{\mu}X=0.
\label{E.O.M.}
\end{equation}

The model can also be defined in a gauge invariant way if we
introduce a projection supermatrix $P$
\begin{equation}
P\equiv
X\,X^{\dagger}=\sum_{j=1}^{p+q} \z_j\z_j^{\dagger},
\label{P-operator}
\end{equation}
which is obviously gauge
invariant under (\ref{Gauge-tr}) and has rank of $p+q$ ($p$
bosonic eigenvectors and $q$ fermionic eigenvectors), and enjoys
the properties,
\begin{equation}
P^{\dagger}=P=P^2.
\end{equation}
The action (\ref{Action}) can be re-expressed in terms of the
projection supermatrix
\begin{equation}
S=\frac{1}{2}\int d^2x\,
\s\lt(\partial_{\mu}P\,\partial_{\mu}P\rt),
\end{equation}
and the corresponding equation of motion becomes
\begin{equation}
\lt[\partial_{\mu}\partial_{\mu}P,\,P\rt]=0.
\label{E.O.M.-1}
\end{equation}

The remaining part $Y$ (\ref{XY}) of the unitary supermatrix $g$
defines another
  projection supermatrix $\bar{P}$:
\begin{equation}
\bar{P}\equiv Y\,Y^{\dagger}=\sum_{j=p+q+1}^{m+n}
\z_j\z_j^{\dagger}=1_{m+n}-P, \label{P-operator-1}
\end{equation}
which is orthogonal to $P$. It has  rank $m+n-(p+q)$ (i.e., $m-p$
bosonic eigenvectors and $n-q$ fermionic eigenvectors). This
projection supermatrix  $\bar{P}$ satisfies the same equation of
motion (\ref{E.O.M.-1}) as that of $P$, reflecting the obvious
symmetry $\{p,q\}\leftrightarrow\{m-p,n-q\}$ of the super
Grassmannian $G(m,p|n,q)\cong U(m|n)/[U(p|q)\otimes U(m-p|n-q)]$.

For the simplest case of $p=1$ and $q=0$, the super Grassmannian
manifold $G(m,p|n,q)$ reduces to the projective superspace ${\bf
CP}^{m-1|n}$ and the  corresponding sigma model was investigated
in \cite{others}.


\section{Exact solutions}
\label{ES} \setcounter{equation}{0}

In this section, we shall construct a series of solutions of the
super Grassmannian sigma model given by the action (\ref{Action}).
These solutions are expressed in terms of a set of holomorphic
bosonic and fermionic supervector input functions.

Let us introduce the complex coordinates of the two-dimensional
Euclidean space
\begin{equation}
x_+=x_1+\i x_2,\quad x_-=x_1-\i
x_2.
\end{equation}
The equation of motion (\ref{E.O.M.}) for $X$ is rewritten as
\begin{equation}
D_+D_-X+X\lt(D_-X\rt)^{\dagger}\lt(D_-X\rt)=0,\quad
\partial_{\pm}=\frac{\partial}{\partial x_{\pm}},
\label{Eq-1}
\end{equation}
or equivalently,
\begin{equation}
D_-D_+X+X\lt(D_+X\rt)^{\dagger}\lt(D_+X\rt)=0.
\label{Eq-2}
\end{equation}
Likewise, one may rewrite the gauge invariant equation
(\ref{E.O.M.-1}) as
\begin{equation}
\lt[\partial_+\partial_-P,\,P\rt]=0. \label{Eq-3}
\end{equation}

To construct generic solutions to the equation of motion, let us
consider as input  $(p+q)$ linearly independent {\em holomorphic}
{\it pure}   supervectors of $(m,n)$ type, among which $p$ are
{\em bosonic} and $q$ are {\em fermionic}, but the order is
completely arbitrary, as mentioned above. Let us denote
them
\begin{equation}
\f_1, \,\f_2, \ldots,
\f_{p+q},\qquad \partial_-\f_j=0,\quad j=1, \ldots,p+q.
\label{inidata}
\end{equation}
A different ordering of these $(p+q)$ input supervectors will give
rise to different solutions.
 From these supervectors $\{\f_j\}$, $j=1,\ldots,p+q$ we construct $m+n$  holomorphic
 pure supervectors  of $(m, n)$ type by successive
differentiation with  respect to $x_+$:
\begin{align}
\f_{p+q+1}&=\partial_+\f_1,\  \f_{p+q+2}=\partial_+\f_2,\
  \ldots, \ \f_{2p+2q}=\partial_+\f_{p+q},\no\\
\f_{2p+2q+1}&=\partial_+^2\f_1,\ \f_{2p+2q+2}=\partial_+^2\f_2, \ldots,
\f_{3p+3q}=\partial_+^2\f_{p+q},\no\\
\f_{3p+3q+1}&=\partial_+^3\f_1,\ \f_{3p+3q+2}=\partial_+^3\f_2,
  \ldots, \f_{4p+4q}=\partial_+^3\f_{p+q},\no\\
&\quad \ldots, \quad \ldots, \quad \f_{m+n}
\label{Constru-2}
\end{align}
Supposing  that the resulting supervectors
$\f_1,\ldots,\f_{m+n}$ are linearly independent and their body
vectors
  $\f_{1\,B}$,\ldots, $\f_{m+n\, B}$ are also linearly independent,
  we apply the Gramm-Schmidt procedure to obtain an
orthonormal basis of the superspace $\V$:
\begin{equation}
\e_1,\ldots,\e_{m+n},\quad \e_j^{\dagger}\e_k=\delta_{j\, k}.
\label{Uni-1}
\end{equation}
To be more specific,
the Gramm-Schmidt procedure goes as follows
\begin{equation}
\begin{array}{ll}
\e_1=\frac{\g_1}{||\g_1||},& \g_1=\f_1,\\
\e_2=\frac{\g_2}{||\g_2||},& \g_2=\f_2-\e_1\,(\e_1^{\dagger}\,\f_2),\\
\quad\,\vdots& \quad\,\vdots\\
\e_j=\frac{\g_j}{||\g_j||},&
\g_j=\f_j-\sum_{k=1}^{j-1}\e_k\,(\e_k^{\dagger}\,\f_j),\\
\quad\,\vdots& \quad\,\vdots
\end{array}
 \label{grschm}
\end{equation}
It is easy to check that the resulting basis vectors $\e_i$ are
pure vectors and the $\Zb_2$-grading has a period
$p+q$ because of the preparation of the supervectors
$\{\f_j\}$ 
(\ref{Constru-2}):
\begin{equation}
[\e_{p+q+j}]=[\e_j], \quad j=1,\ldots,m+n-(p+q).
\end{equation}

Here is one important remark concerning the Gramm-Schmidt
orthonormalisation of the above supervectors (\ref{Constru-2}).
For the generic case of the chosen integers $m$, $n$, $p$ and $q$,
the above Gramm-Schmidt procedure does not come to the end
$\e_{m+n}$ but it stops at  $\e_N$,  for certain $N\leq m+n$.
There are $m$ even and $n$ odd basis vectors in $\V$. However,
the orthonormalisation proceeds by the unit of $p$ even and $q$
odd supervectors and the choice of $p$ and $q$ is independent of
$m$ and $n$ except for the obvious constraints $0\leq p\leq m$,
$0\leq q\leq n$ and $1\leq p+q$. Therefore it can generically
happen that either the entire $m$ even basis or the entire $n$ odd
basis is already made before all the vectors in (\ref{inidata}),
(\ref{Constru-2}) can be orthonormalised by means of
(\ref{grschm}) to end with $\e_{m+n}$. Let $\f_{N+1}$ be the
$(m+1)$-th even  supervector or the $(n+1)$-th odd supervector in
(\ref{Constru-2}) to be orthonormalised. Then its projection
\begin{equation}
\g_{N+1}=\left(1_{m+n}-\sum_{j=1}^N\e_j\e_j^\dagger\right)
\f_{N+1}
\end{equation}
cannot have a non-vanishing body and
therefore cannot be normalised. If it has, $\g_{N+1}$ can be
normalised to obtain the $(m+1)$-th even base or $(n+1)$-th odd
base, which cannot happen in a supervector space $\V$ of $(m,n)$
type. To summarise, the set of orthonormal supervectors obtained
by the above procedure (\ref{grschm}) will be
\begin{equation}
\left\{\e_j\mid\ j=1,2,\ldots,\mbox{min}(N,m+n)\right\}.
\end{equation}
Hereafter we use the notation $N$ as meaning min$(N,m+n)$ for
simplicity.

At first sight this might seem rather strange. But at closer
inspection, it turns out to be rather natural. For instance,
consider the extreme case of $q=0$. In this case we start with
only the bosonic input supervectors. The orthonormalisation
produces the bosonic base vectors only, and it stops at $\e_m$.
This phenomenon of intermediate stopping of the orthonormalisation
procedure does not happen in the non-super Grassmannian sigma
models \cite{Sas83}.

\bigskip
By picking up $p+q$ {\em consecutive orthonormal supervectors\/},
we define the following $(m+n)\times(p+q)$ matrices:
\begin{equation}
\begin{array}{rl}
X_{(1)}&=(\e_1,\e_2,\ldots,\e_{p+q}),\\
X_{(2)}&=(\e_2,\e_3,\ldots,\e_{p+q+1}),\\
\vdots\quad&\qquad\quad \vdots\\
X_{(N-p-q+1)}&=(\e_{N-p-q+1},\ldots,\e_{N}).
\end{array}
\label{sols}
\end{equation}

All  satisfy the constraint (\ref{Orthnormal}) and they contain
$p$ even and $q$ odd basis supervectors. But the order  of the
even and the odd basis is not the same.  Then $X_{(j)}$, $j=1$,
\ldots, $N-p-q+1$
  and their gauge transformed form $X'_{(j)}=X_{(j)}h_{(j)}$
{\em satisfy the equation of motion\/} (\ref{Eq-1}).
The proof is quite elementary.

Let us start the proof with the analytic property of the
orthonormal supervectors. Because of the orthonormalisation
procedure, we know that each basis vector $\e_i$ can be expressed
in terms of $\{\f_j\}$, $j=1$,\ldots, $i$:
\begin{equation}
\e_i=\sum_{j=1}^i\f_j\,a_j,\qquad i=1,\ldots,N,
\label{Expansion-1}
\end{equation}
where the coefficients
$\{a_j\}$ are supernumbers. The above expression implies that the
expansion of each vector $\f_i$ in terms of the basis $\{\e_i\}$
has the same triangular form
\begin{equation}
\f_i=\sum_{j=1}^i\e_j\,b_j,\qquad i=1,\ldots,N,
\label{Expansion-2}
\end{equation}
with some supernumbers
$\{b_j\}$. By differentiating (\ref{Expansion-1}) with respect to
$x_-$, we have $\partial_-\e_i=\sum_{j=1}^i\f_j\,\partial_-a_j$
since  $\{\f_j\}$ are holomorphic supervectors. Using the
expansion (\ref{Expansion-2}), we find that
\begin{equation}
\partial_-\e_i=\sum_{j=1}^i\e_j\,(\e^{\dagger}_j\,\partial_-\e_i),
\quad i=1,\ldots,N. \label{3.12}
\end{equation}
Moreover, by the ways of constructing the vectors
$\{\f_1,\ldots,\f_{m+n}\}$
(\ref{Constru-2}) we have the following useful expansions
\begin{equation}
\partial_+\e_i=\sum_{j=1}^{i+p+q}\e_j\,(\e^{\dagger}_j\partial_+\e_i),
\qquad 1\leq i\leq N-(p+q). \label{3.18a}
\end{equation}
The above two relations (\ref{3.12})-(\ref{3.18a}) will play an
essential role in the proof  for the solutions (\ref{sols}).

We will show that
\begin{equation}
X_{(j)}=(\e_j,\e_{j+1},\ldots,\e_{j+p+q-1}), \quad
j=1,2,\ldots,N-(p+q)+1,
\end{equation}
solves the equation of motion (\ref{Eq-1}), or the corresponding
projection supermatrix
\begin{equation}
P_{(j)}=X_{(j)}\,X_{(j)}^\dagger=\sum_{k=j}^{j+p+q-1}\e_k\e_k^\dagger,
\quad P_{(j)}^\dagger=P_{(j)}=P_{(j)}^2,
\label{jproj}
\end{equation}
satisfies (\ref{Eq-3}):
\begin{equation}
[\partial_+\partial_-P_{(j)},P_{(j)}]=0.
\end{equation}
Although proper care is needed for the grading problem, most
formulas formally look essentially the same as those in the
non-super case \cite{Sas83}. We will proceed in a similar way as
those in
the non-super case. Following \cite{Sas83}, we introduce an
auxiliary supermatrix variable $Q_{(j)}$ by
\begin{equation}
Q_{(j)}=\sum_{k=1}^{j-1}\e_k\e_k^\dagger,\qquad
j=1,2,\ldots,N-(p+q)+1,
\end{equation}
which is a projection supermatrix, too:
\begin{equation}
Q_{(j)}^\dagger=Q_{(j)}=Q_{(j)}^2.
\end{equation}
It is of rank $j-1$ and is orthogonal to $P_{(j)}$
\begin{equation}
P_{(j)}Q_{(j)}=Q_{(j)}P_{(j)}=0.
\label{pqortho}
\end{equation}
Hereafter the suffix $(j)$ of $P_{(j)}$ and $Q_{(j)}$ is fixed and
will not be written explicitly to make the notation simple. {}From
the $x_-$ derivative relation (\ref{3.12}) we obtain
\begin{equation}
(\partial_-Q)Q=0,
\label{dminQ}
\end{equation}
and
\begin{equation}
\partial_-(P+Q)(P+Q)=0.
\label{dminPQ}
\end{equation}
Taking the $x_-$ derivative of the orthogonality relation
(\ref{pqortho}), we obtain
\begin{equation}
(\partial_-P)Q+P(\partial_-Q)=0.
\label{dminortho}
\end{equation}
Another simple consequence of the $x_-$ derivative relation
(\ref{3.12}) is
\begin{equation}
P(\partial_-Q)=0. \label{PdminQ}
\end{equation}
By combining (\ref{dminQ})--(\ref{PdminQ}), we obtain a simple
relationship
\begin{equation}
(\partial_-P)P+(\partial_-Q)P=0.
\label{dminpp}
\end{equation}

Next we consider the consequences of the $x_+$ derivative relation
(\ref{3.18a}). The only  essential  formula  is
\begin{equation}
P(\partial_+Q)=\partial_+Q.
\label{Relation-1}
\end{equation}
This reflects the facts that the $x_+$ differentiation sends
$\f_k$ to $\f_{k+p+q}$ (\ref{Constru-2}) and that the  projection
supermatrix $P$ consists of just the $p+q$ orthonormal
supervectors after $Q$. We will give a simple proof in the
Appendix. The hermitian conjugation of the above formula reads
\begin{equation}
(\partial_-Q)P=\partial_-Q.
\label{dminqp}
\end{equation}
Equation (\ref{dminpp}) and (\ref{dminqp}) combine to give
\begin{equation}
(\partial_-P)P+\partial_-Q=0
\label{dminfin}
\end{equation}
together with its hermitian conjugation
\begin{equation}
P(\partial_+P)+\partial_+Q=0.
\label{dplufin}
\end{equation}
By subtracting the $x_-$ derivative of (\ref{dplufin}) from the
$x_+$ derivative of (\ref{dminfin}),  we obtain the desired formula
\begin{equation}
[\partial_+\partial_-P,P]=0,
\end{equation}
which completes the proof.


\section{Comments and Summary}
  \label{Con}
\setcounter{equation}{0}

Some comments and remarks are in order. The first is about {\em
instanton\/} and {\em anti-instanton solutions\/}. As is obvious
from (\ref{Eq-1}) and  (\ref{Eq-2}), if $X$ satisfies
\begin{equation}
D_-X=0,\qquad \mbox{or}\qquad D_+X=0
\label{firstorder}
\end{equation}
then the full equation of motion
is trivially satisfied. These are simply the covariant version of
the equations characterising the holomorphic and anti-holomorphic
functions
\[
\partial_-f=0,\qquad \mbox{or}\qquad \partial_+f=0.
\]
In analogy with the solutions of the (anti-)self-dual equations
for the gauge field strength, which automatically satisfy the full
second order equations, these solutions are called  {\em
instanton\/} and {\em anti-instanton solutions\/}, respectively.
In terms of the projector supermatrix $P$, the first order
equations (\ref{firstorder}) are written as
\begin{equation}
(\partial_-P)P=0,\qquad \mbox{or}\qquad P(\partial_-P)=0.
\label{pinst}
\end{equation}
By differentiating the first equation with respect to $x_+$, we
obtain
\begin{equation}
(\partial_+\partial_-P)P+(\partial_-P)(\partial_+P)=0,
\end{equation}
and its hermitian conjugation
\begin{equation}
P(\partial_+\partial_-P)+(\partial_-P)(\partial_+P)=0.
\end{equation}
By subtracting these two equations, we obtain
$[\partial_+\partial_-P,P]=0$. Thus the full equation of motion
follows if either of (\ref{pinst}) is  satisfied. Among our
explicit solutions, the first one, $X_{(1)}$, which is obtained from
the input supervectors only without any differentiation, is the
instanton solution. For $j=1$, $Q_{(j)}=0$ and (\ref{dminpp}),
$(\partial_-P)P+(\partial_-Q)P=0$, simply means
$(\partial_-P_{(1)})P_{(1)}=0$. As is easily expected the
anti-instanton is the last one, $X_{(m+n-p-q+1)}$. For
$j=m+n-p-q+1$, one has $Q_{(j)}+P_{(j)}=1_{m+n}$. Thus
(\ref{PdminQ}), $P(\partial_-Q)=0$ means
$P_{(j)}(\partial_-P_{(j)})=0$. As remarked earlier, our
orthonormalisation procedure might not come to the end and the
anti-instanton might not be included in our set of solutions. It
may seem that our solution generation method discriminates the
anti-instantons over instantons but the situation could be
reversed if we decide to use the  anti-holomorphic supervectors as
input.

A few words on other types of solutions. The solutions explored in
section \ref{ES} are called {\em generic\/}, since they depend on
the maximal number of input data for $G(m,p|n,q)$, $p$ bosonic and
$q$ fermionic holomorphic supervectors. A closer look at the proof
might reveal that the same construction method with less input
holomorphic bosonic (fermionic) supervectors also produces
solutions of the super Grassmannian model $G(m,p|n,q)$. They are
called degenerate solutions after the non-super case \cite{Sas83}.
Furthermore, one may also construct some reducible solutions in
the sense of \cite{Sas83} from these resulting degenerate ones as
those in the non-super case. The completeness of the solutions
thus obtained is beyond the scope of the present paper.

As is well known, any solution of the super Grassmannian
$G(m,p|n,q)\cong U(m|n)/[U(p|q)\otimes U(m-p|n-q)]$ sigma model
for various $p$ and $q$ provides a very special class of solutions
of the supergroup (or the principal chiral) $U(m|n)$ sigma model.
Take \[ g=1_{m+n}-2P_{(j)},\qquad j=1,\ldots,N-(p+q)+1, \] for any
$j$ and for arbitrary $p$ and $q$. It is a special element of
$U(m|n)$ satisfying the condition $g^2=1_{m+n}$ and the  equation
of motion:
\begin{equation}
\partial_\mu(g^{-1}\partial_\mu g)
=-2\left[\vTs \partial_\mu\partial_\mu P_{(j)},P_{(j)}\right]=0,
\end{equation}
thanks to the projection properties of $P_{(j)}$ (\ref{jproj}).

Here is a summary. We have formulated non-linear sigma models on
certain supermanifolds, the complex super Grassmannian
$G(m,p|n,q)$ including the super projective spaces ${\bf
CP}^{m-1|n}$. The base space is the two-dimensional Euclidean
space. These are massless scalar field theories with non-linear
geometrical constraints due to the supermanifolds. A wide class of
classical exact solutions, or  {\em harmonic maps\/} are
constructed explicitly and elementarily in terms of the
Gramm-Schmidt orthonormalisation procedure starting from the input
holomorphic supervectors of type $(m,n)$, among them $p$ bosonic
and $q$ fermionic.

\section*{Acknowledgements}
The financial support from  Australian Research Council  is
gratefully acknowledged. R.S. is supported in part by Grant-in-Aid
for Scientific Research from the Ministry of Education, Culture,
Sports, Science and Technology, No.18340061 and No.16340040.
Y.Z.Z. would like to thank for hospitality  Yukawa Institute for
Theoretical Physics, Kyoto University, and Condensed Matter Theory
Laboratory, RIKEN Discovery Research Institute, where part of  the
work was done. R.S. thanks Department of Mathematics, University
of Queensland for hospitality.


%
\section*{Appendix:  Proof of (\ref{Relation-1}) } \setcounter{equation}{0} \renewcommand{\theequation}{A.\arabic{equation}}

Here we provide a straightforward proof of (\ref{Relation-1}),
which is essential for the construction of solutions. Using the
relation (\ref{3.12})-(\ref{3.18a}), we evaluate $\partial_+Q$:
\begin{eqnarray}
  \partial_+Q&=&\sum_{l=1}^{j-1}
\lt((\partial_+\e_l)\,\e^{\dagger}_l
+\e_l\,\partial_+\e^{\dagger}_l\rt) = \sum_{l=1}^{j-1}
\lt((\partial_+\e_l)\,\e^{\dagger}_l+
\e_l\,\lt(\partial_-\e_l\rt)^{\dagger}\rt)
\nonumber\\
&\stackrel{(\ref{3.12})}{=}& \sum_{l=1}^{j-1}\left[
(\partial_+\e_l)\,\e^{\dagger}_l+\e_l
\sum_{k=1}^l\lt(\e_k\,\e_k^{\dagger}\partial_-\e_l\rt)^{\dagger}\right]
\nonumber\\
&\stackrel{(\ref{3.18a})}{=}&
\sum_{l=1}^{j-1}\left[\sum_{k=1}^{l+p+q}
\e_k\lt(\e^{\dagger}_k\partial_+\e_l\rt)\e^{\dagger}_l
+\e_l\sum_{k=1}^l
\lt(\partial_+\e^{\dagger}_l\,\e_k\rt)\e^{\dagger}_k\right]\nonumber\\
&=&\sum_{l=1}^{j-1}\sum_{k=1}^{l+p+q}
\e_k\lt(\e^{\dagger}_k\partial_+\e_l\rt)\e^{\dagger}_l
-\sum_{k=1}^{j-1}\sum_{l=1}^k\e_k
\lt(\e^{\dagger}_k\,\partial_+\e_l\rt)\e^{\dagger}_l.
\end{eqnarray}
Among the first summation terms, we decompose \[
\sum_{k=1}^{l+p+q}\e_k =\sum_{k=1}^{j-1}\e_k
+\sum_{k=j}^{l+p+q}\e_k.
\]
The second sum is annihilated if multiplied by $(1-P)$ from the
left and we obtain \begin{align}
(1-P)\partial_+Q&=(1-P)\left[\sum_{l=1}^{j-1}
\sum_{k=1}^{j-1}-\sum_{k=1}^{j-1}\sum_{l=1}^k\right]
\e_k\lt(\e^{\dagger}_k\partial_+\e_l\rt)\e^{\dagger}_l
\nonumber\\
&=(1-P)\sum_{k=1}^{j-1}\sum_{l=k+1}^{j-1}\,
\e_k\lt(\e^{\dagger}_k\partial_+\e_l\rt)\e^{\dagger}_l
\nonumber\\
&=-(1-P)\sum_{k=1}^{j-1}\sum_{l=k+1}^{j-1}\,
\e_k\lt((\partial_-\e_k)^\dagger \e_l\right)\e_l^\dagger=0.
\end{align}
The last equality is due to (\ref{3.12}). Thus we obtain
$P(\partial_+Q)=\partial_+Q$ and   (\ref{Relation-1}) is proved.


\end{document}